\documentclass{article}

\usepackage{arxiv}

\usepackage[utf8]{inputenc}
\usepackage[T1]{fontenc}
\usepackage{url}
\usepackage{booktabs}
\usepackage{amsfonts}
\usepackage{amsmath}
\usepackage{amssymb}
\usepackage{bm}
\usepackage{nicefrac}
\usepackage{microtype}
\usepackage{graphicx}
\usepackage{subcaption}
\usepackage{siunitx}
\usepackage{xcolor}
\usepackage[section]{placeins}
\usepackage{hyperref}  
\graphicspath{{./figures/}{./images/}}

\newcommand{\x}{\bm{x}}
\newcommand{\Ug}{U_G}
\newcommand{\Ul}{U_{\ell}}
\newcommand{\Dn}{\Delta n}
\newcommand{\We}{\mathrm{We}}
\newcommand{\Rel}{\mathrm{Re}_{\ell}}

\newcommand{\rmsephi}{\mathrm{RMSE}_{\phi}}
\newcommand{\rmsealphap}{\mathrm{RMSE}_{\alpha^p}}
\newcommand{\rmsealpha}{\mathrm{RMSE}_{\alpha}}
\newcommand{\epsm}{\varepsilon_m}
\newcommand{\reg}{\delta}                       
\newcommand{\Dtrain}{\mathcal{D}_{\mathrm{train}}}
\newcommand{\Drange}{\mathcal{D}_{\mathrm{range}}}
\newcommand{\Dtest}{\mathcal{D}_{\mathrm{test}}}

\newcommand{\FNOsdf}{\mathrm{FNO}_{\phi}}
\newcommand{\FNOsdfPI}{\mathrm{FNO}_{\phi}^{\mathrm{PI}}}
\newcommand{\FNOalpha}{\mathrm{FNO}_{\alpha}}
\newcommand{\UNetalpha}{\mathrm{U\text{-}Net}_{\alpha}}

\title{Towards Rapid Prototyping of Spray Injectors: \\
A Regime-Agnostic Neural-Operator Surrogate for
Gas-Liquid Interface Evolution}
\author{
Paolo Guida$^{*}$ \\
Clean Energy Research Center \\
King Abdullah University of Science and Technology \\
Thuwal 23955, Saudi Arabia \\
\texttt{paolo.guida@kaust.edu.sa} \\
\And
Po-Han Chen \\
Clean Energy Research Center \\
King Abdullah University of Science and Technology \\
Thuwal 23955, Saudi Arabia \\
\texttt{pohan.chen@kaust.edu.sa} \\
\And
Hong G. Im \\
Clean Energy Research Center \\
King Abdullah University of Science and Technology \\
Thuwal 23955, Saudi Arabia \\
\texttt{hong.im@kaust.edu.sa} \\
\And
William L. Roberts \\
Clean Energy Research Center \\
King Abdullah University of Science and Technology \\
Thuwal 23955, Saudi Arabia \\
\texttt{william.roberts@kaust.edu.sa} \\
}
\begin{document}

\maketitle

\renewcommand{\thefootnote}{\fnsymbol{footnote}}
\footnotetext[1]{Corresponding author: \texttt{paolo.guida@kaust.edu.sa}}
\renewcommand{\thefootnote}{\arabic{footnote}}

\begin{abstract}
Spray atomisation is used in a variety of applications that rely on its ability to instantaneously create an extremely large surface area between the liquid and gas phases. However, predictive analysis of spray behaviour and estimates of surface area are extremely complex. Experimental activities are constrained by diagnostics that cannot access all spray regions, while numerical methods are computationally expensive, particularly as finer structures form. Data-driven methods can address this issue by learning how interfaces are generated and evolve in space, enabling users to replace complex, often slow CFD simulations with fast inference, thereby quickly exploring design space, ranking conditions, and eventually finely controlling spray atomisation.
In this work, we propose an architecture that learns the most relevant parameter in sprays: their liquid-gas surface evolution. In doing so, we evaluate the effects of state representation, neural architecture, and physics-informed regularisation on autoregressive forecasting of spray interfaces, with particular attention to conservation behaviour over long horizons. The principal method we use is a boundary-conditioned Fourier Neural Operator that learns the evolution of the Signed Distance Function (SDF) from the gas-liquid interface. We trained the model on a dataset spanning several atomisation regimes. The dataset consists of 2D CFD simulations performed using the VoF sharp-interface method. We evaluated the baseline SDF-based FNO surrogate against the same prediction made with U-Net and with an FNO trained on the volume fraction itself.
We observed that the proposed SDF-FNO model retains better fidelity than the FNO trained on the volume fraction but it is outperformed by the U-Net architecture. An ablation of the training objective shows that the liquid-inventory penalty buys conservation, at a modest cost to local interface fidelity.
We then introduce a physics-informed extension that combines a target-increment liquid-balance penalty for the open domain, a narrowband Eikonal regulariser that preserves signed-distance geometry, and a phase-boundedness penalty, and we evaluate it directly against the data-driven baseline, finding that it optimises stably but leaves error, interface overlap, and inventory behaviour essentially unchanged.
We finally show a practical application of the developed surrogate, using it to rank injection conditions by the interfacial area generated per unit gas-injection power across the operating envelope of a fixed geometry.
\end{abstract}

\section{Introduction}
Atomising sprays govern the performance of a wide range of chemical process equipment, from fuel injectors and spray dryers to scrubbers, absorbers, and coating units \cite{lefebvre2017atomization,rigas2016spray,moghimi2025simulations}.
All of these depend on breaking a liquid stream into droplets, and the interfacial area generated by that breakup sets the available rate of interphase heat and mass transfer, or the deposition rate \cite{ishii2010thermo,lefebvre2017atomization}. That interfacial area is produced by a fast, multiscale process: high-speed multiphase flows are governed by interfaces that stretch, fragment, reconnect, and are transported across a wide range of spatial and temporal scales. In this scenario, explicit gas-liquid interface tracking becomes progressively more computationally expensive as the flow develops.

Designing an injector, therefore, involves searching an operating envelope for the required spray topology \cite{xia2020experimental,di2022computational,sun2024liquid}. That search is currently either fully experimental, correlation-based, or reliant on computational fluid dynamics simulation. Experimental campaigns are time-consuming, and spray diagnostics often cannot capture the complete transient three-dimensional interface or provide sufficient detail to quantify its evolution \cite{fansler2015spray}. The correlation route is fast, but correlations generally derive from experiments, and collapsing the spray to a mean droplet diameter discards the spatial structure on which the design often depends \cite{lefebvre2017atomization,dong2023development}. Interface-resolved computational fluid dynamics (CFD) has the potential to retain that structure and reliably resolve the underlying behaviour \cite{desjardins2008accurate,shrestha2023primary,sun2024liquid}, but at a cost that renders it unusable for wide-scale design exploration, optimisation, or closed-loop control. Existing high-fidelity strategies, in fact, each trade one difficulty for another. While Lagrangian frameworks, by tracking the evolution and trajectories of each droplet, enable agile computations, they also require extensive parameter tuning and are not generally suitable for extrapolating behaviours. Interface-capturing methods, on the other hand, are, in theory, predictive, but guaranteeing enough resolution to resolve the droplets while simulating a real-size device is exceptionally demanding \cite{desjardins2008accurate,GAMET2020104722}. Phase-field and hybrid formulations offer further compromises \cite{heinrich20203d,qiu2022physics}. Data-driven methods offer a potential solution to these limitations \cite{salehi2025data,azizzadenesheli2024neural}. A surrogate model that predicts the full interface field across operating conditions at a fraction of the computational cost can bridge the gap between empirical correlations and detailed simulations \cite{guida2026data,ramlau2026learning}. These surrogates are expected to resolve the sharp interface while respecting constraints like mass and energy balance \cite{guida2026data,lin2026operator}.

Neural operators are architectures designed to learn maps between function spaces, enabling them to approximate a PDE solver across a range of operating conditions. Among the various proposed frameworks, Fourier Neural Operators (FNOs) have been increasingly adopted in the field of fluid dynamics for their ability to map through convolution in spectral space~\cite{li2020fourier,tran2021factorized,NEURIPS2023_940a7634,duruisseaux2025fourier}. They have been applied to two-phase, subsurface, and reaction-diffusion systems~\cite{zhang2022fourier,jain2025scaling,ma2024enhancing,dong2025multi,hao2024fourier,lyu2023multi}, and physics-informed operator variants have been developed for two-phase and multi-phase-field settings~\cite{goswami2023physics,qiu2022physics,zhang2024physics}.

In this work, we propose adopting FNOs to predict spray evolution and training them on a dataset built from detailed sharp-interface CFD simulations spanning several spray regimes within a single injector geometry. Our goal is to identify an architecture capable of predicting the interface's evolution over time and the most significant parameters for evaluating a spray. In particular, in this work, we focus on the liquid volume fraction, which is the standard output retrieved from Volume of Fluid computations. While the volume fraction is an appropriate parameter to track, it does pose some issues. In particular, for the cases explored in this work, it is important to maintain a sharp interface. By a sharp interface, we mean keeping the interface within a single computational cell. This results in very abrupt gradients that may not be compatible with an FNO description. Direct regression of the phase fraction is, in fact, difficult because the target is discontinuous across the interface, and pixel-wise regression losses poorly serve sharp transitions. We therefore propose replacing the phase fraction with a signed-distance representation of the same interface, which turns a discontinuous target into a smooth, monotonic geometric field,
\begin{equation}
    \phi(\x,t)
    =
    \begin{cases}
        -d(\x,\Gamma_t), & \x\in\Omega_{\ell}(t),\\
        +d(\x,\Gamma_t), & \x\notin\Omega_{\ell}(t),
    \end{cases}
    \label{eq:sdf_definition}
\end{equation}
where $\Gamma_t$ is the liquid interface, $\Omega_{\ell}(t)$ the liquid region, and $d(\x,\Gamma_t)$ the shortest distance from $\x$ to the interface. The zero level set of $\phi$ recovers the interface. At the same time, its magnitude measures the distance from it, so that forecasting the discontinuous phase field is recast as forecasting a smoother geometric field. Related surrogate work has similarly found that geometry-aware representations of the interface are more learnable than the raw phase state~\cite{ramlau2026learning,guida2026data}.

This is why we explore three factors: the field representation the operator predicts, the neural architecture that performs the prediction, and the effect of physics-informed constraints applied during training.

The core model we propose is a boundary-conditioned FNO that predicts a signed-distance field and reconstructs a volume-fraction proxy via a differentiable transformation. The model is trained autoregressively on 50 different conditions, each with a combination of liquid and gas inlet velocities. Against this, we test two direct-phase comparators, an FNO and a U-Net~\cite{ronneberger2015u}, and, within the signed-distance model, a physics-regularised extension combining a target-increment liquid-balance penalty, a narrowband Eikonal term, and a phase-boundedness constraint~\cite{raissi2019physics}.
The work is structured as follows:
\begin{itemize}
    \item We first isolate the effect of the state representation on spectral operator learning, achieving an error reduction in the autoregressive rollout of the FNO trained on the signed distance function with respect to that trained on the volume fraction directly. The two models are scored in their respective target spaces. Hence, the comparison measures the benefit of predicting a geometric field and reconstructing the phase from it, rather than a like-for-like error ratio.
    \item We then compare the FNO with a different architecture commonly used in this type of application, the U-Net, using identical training and evaluation protocols to highlight the strengths and weaknesses of both approaches.
    \item We conduct an ablation study of the signed-distance model that exposes a trade-off between geometric fidelity and liquid-inventory preservation, and we introduce a physics-regularised neural operator architecture to evaluate the effect of adding further constraints to the model. The ablation is what establishes the value of inventory regularisation here; the physics-informed extension, reported in Section~\ref{sec:physics_informed}, optimises stably but does not improve prediction.
    \item Finally, we propose an application of the surrogate model to injector screening, in which we use the learned model to build atomisation-onset boundaries, constrained efficiency maps, and a feasible area-power Pareto front. We establish that the surrogate, while biased, can still serve as a rank-preserving interfacial-area estimator, thereby making it usable for screening the operating envelope of the geometry on which it was trained.
\end{itemize}
Lastly, we report the current limitations of our approach and potential future development opportunities.

\section{Problem formulation and dataset}

We built the dataset by performing numerical simulations with a two-dimensional spray solver based on the volume-of-fluid method~\cite{hirt1981volume} in OpenFOAM. We treated interface advection geometrically using the \texttt{isoAdvector} algorithm~\cite{roenby2017new}, combined with the \texttt{plicRDF} algorithm~\cite{scheufler2019accurate} for geometric interface reconstruction. We obtained the interface normal from the reconstructed distance function~\cite{cummins2005estimating}, which we also used to evaluate the interface curvature~\cite{guida2022computational,scheufler2023twophaseflow} required for the surface-tension force. Because surface tension plays a central role in spray breakup, we considered accurate interface reconstruction and curvature evaluation essential to the fidelity of our simulations. The accuracy of this numerical framework has been demonstrated through several benchmarks~\cite{guida2022computational,gamet2020validation} and through successful validation against experimental measurements of coaxial airblast atomisation~\cite{chen2026large,chen2026effects}. The properties of the fluids simulated are reported in Table~\ref{tab:properties}.

\begin{table}[ht!] \footnotesize
\caption{Fluid properties used in the simulations of coaxial airblast atomisation for dataset generation.}
\centerline{\begin{tabular}{lccc}
\hline
& Density & Kinematic viscosity & Surface tension\\
& \si{\kilo\gram\per\cubic\metre} & $\times10^{-6}$ \si{\square\metre\per\second} & $\times10^{-3}$ \si{\newton\per\metre} \\
\hline
Ethanol & 790 & 1.51 & 22.1 \\
Air & 1.204 & 15.1  & - \\
\hline
\end{tabular}}
\label{tab:properties}
\end{table}

\begin{figure}
    \centering
    \includegraphics[width=0.5\linewidth]{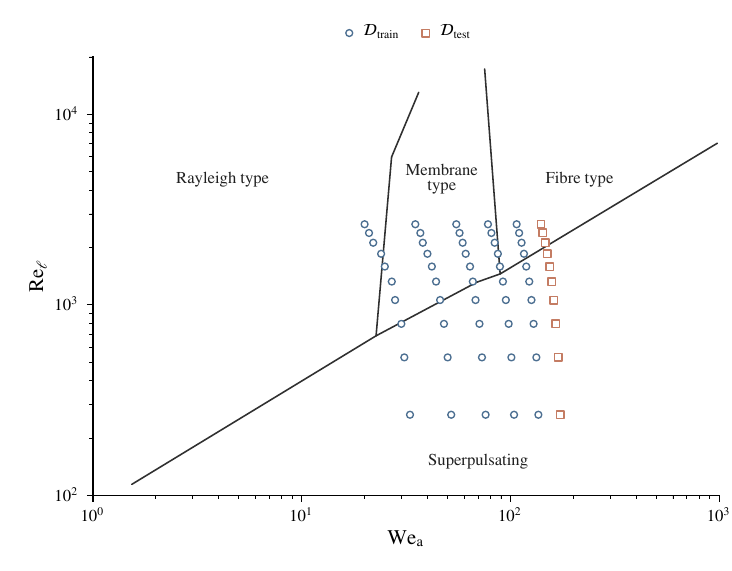}
    \caption{%
        Distribution of the dataset over the atomisation-regime map, spanned by the aerodynamic Weber number $\We_a$ and the liquid Reynolds number $\Rel$~\cite{lefebvre2017atomization}. Training conditions ($\Dtrain$, circles) tile the interior of the map, covering the membrane, fibre, and superpulsating regimes, while the extrapolation conditions ($\Dtest$, squares) lie along the high-$\We_a$ edge, deeper into the fibre-type regime than any training case. The solid lines mark the approximate boundaries between atomisation regimes.}
    \label{fig:regime_map}
\end{figure}

In our dataset, each case represents a spray driven by a coaxial gas stream, so the fields capture the interaction between the injected liquid and the surrounding high-speed gas. Because our goal is to develop a single operator that predicts spray dynamics across the entire operating envelope of one geometry, rather than a model tuned to a single condition, we span a broad range of gas and liquid velocities. In Figure~\ref{fig:regime_map}, we report a regime map summarising the conditions covered. We note that the simulations are two-dimensional, so the regime map serves to document the coverage of the operating envelope rather than to claim physical equivalence with the three-dimensional experiments from which the regime boundaries derive; in particular, azimuthal instabilities and Plateau-Rayleigh ligament breakup are not represented, and atomisation should be read here as two-dimensional fragmentation dynamics.

We generated 50 operating conditions by combining five gas velocities, $\Ug\in\{40,50,60,70,80\}$, with ten liquid velocities, $\Ul\in\{1,\ldots,10\}$. Each simulation contains $N_t=401$ snapshots on a structured $N_y\times N_x=500\times100$ grid. We reserved ten additional cases at $\Ug=90$ exclusively for high-velocity extrapolation tests. For each condition $c$, identified by $\bm{\mu}^{(c)}=(\Ug^{(c)},\Ul^{(c)})$, we store
\begin{equation}
    \mathcal{D}^{(c)}
    =
    \left\{
        \phi_n^{(c)}(\x),
        \alpha_n^{(c)}(\x)
    \right\}_{n=0}^{N_t-1},
\end{equation}
where $\alpha$ is the phase fraction obtained from CFD and $\phi$ is the corresponding signed-distance field (SDF). We train the network on $\phi$ and reconstruct a phase proxy $\alpha^p$ whenever a phase-fraction representation is required. In this work the hat denotes a prediction, whereas the superscript $\mathrm{GT}$ denotes the ground truth. We normalise the SDF using statistics computed exclusively from the 50 training conditions:
\begin{equation}
    \bar{\phi}_n
    =
    \frac{\phi_n-\mu_{\phi}}{\sigma_{\phi}},
    \qquad
    \mu_{\phi}=4.5923\times10^{-3},
    \qquad
    \sigma_{\phi}=7.9766\times10^{-3},
    \label{eq:sdf_normalisation}
\end{equation}
and map each operating parameter to $[-1,1]$ according to
\begin{equation}
    \bar{U}
    =
    2\frac{U-U_{\min}}{U_{\max}-U_{\min}}-1.
    \label{eq:parameter_normalisation}
\end{equation}
By explicitly supplying the normalised operating parameters, we condition a single operator across the entire training range rather than learning an independent model for each condition.

We construct training samples using a sliding window containing $K=4$ history frames and $R$ target frames. Given the most recent $K$ normalised SDFs, the operator predicts the next field as
\begin{equation}
    \widehat{\bar{\phi}}_{n+1}
    =
    \mathcal{G}_{\theta}
    \left(
        \bar{\phi}_{n-K+1:n},
        \bar{\bm{\mu}},
        \bm{s}
    \right),
    \label{eq:one_step_operator}
\end{equation}
where $\bm{s}$ contains the coordinate and static-geometry channels. We obtain multi-step forecasts by recursively feeding each prediction back into the model:
\begin{equation}
    \widehat{\bar{\phi}}_{n+r}
    =
    \mathcal{G}_{\theta}
    \left(
        \psi_{n+r-K:n+r-1},
        \bar{\bm{\mu}},
        \bm{s}
    \right),
    \qquad r=1,\ldots,R,
    \label{eq:rollout}
\end{equation}
with
\begin{equation}
    \psi_j
    =
    \begin{cases}
        \bar{\phi}_j, & j\leq n,\\
        \widehat{\bar{\phi}}_j, & j>n.
    \end{cases}
\end{equation}
We use $R$ for the relatively short rollout employed during optimisation and $\Dn$ for the generally longer forecast horizon used during evaluation. Mini-batches combine windows from different operating conditions and time indices, allowing each optimisation step to sample multiple regions of the operating map.

\paragraph{Evaluated partitions} It is worth being explicit about what the $60$ evaluated cases are, because the two partitions are not equivalent. The $50$ conditions of $\Drange$ are the training conditions themselves: the operator has seen those velocity pairs during optimisation, and only the specific forecast windows used at evaluation are new. The $10$ conditions of $\Dtest$ at $\Ug=90$ are the sole case-level hold-out. Every aggregate metric reported below is averaged over all $60$ cases and is therefore predominantly in-sample. The only place the two partitions are resolved separately is Figure~\ref{fig:area_validation}, where in-sample and held-out conditions carry distinct markers; readers should keep the composition of the average in mind everywhere else.

\begin{table}[!t]
    \centering
    \caption{Dataset partitions, baseline signed-distance FNO configuration, and training settings.}
    \label{tab:configuration}
    \begin{tabular}{ll}
        \toprule
        Quantity & Value \\
        \midrule
        Training conditions, $|\Dtrain|$
            & $50$ \quad ($\Ug\in\{40,\ldots,80\}\times\Ul\in\{1,\ldots,10\}$) \\
        In-sample evaluated conditions, $|\Drange|$
            & $50$ \quad (the training conditions, evaluated on unseen forecast windows) \\
        Held-out conditions, $|\Dtest|$
            & $10$ \quad ($\Ug=90$, excluded from training) \\
        Total evaluated cases, $|\Drange|+|\Dtest|$ & $60$ \\
        \midrule
        Snapshots per complete case, $N_t$ & $401$ \\
        Spatial resolution ($N_y\times N_x$) & $500\times100$ \\
        History length, $K$ & $4$ \\
        Dynamic input channels & $4$ \\
        Static input channels & $8$ \\
        Total input channels & $12$ \\
        FNO latent width, $d$ & $32$ \\
        FNO depth & $3$ \\
        Retained modes, $(N_{k_y},N_{k_x})$ & $(16,8)$ \\
        Evaluation horizons & $\Dn=1,5,10$ \\
        \bottomrule
    \end{tabular}
\end{table}

\section{Models and training methodology}
\label{sec:models}

For the signed-distance model, we combine the $K$ most recent normalised SDFs with eight conditioning channels:
\begin{equation}
\begin{split}
    \bm{z}_n
    =
    [
        &\bar{\phi}_{n-K+1},
        \ldots,
        \bar{\phi}_{n},
        \bar{x},
        \bar{y},
        \bar{U}_g,
        \bar{U}_{\ell},\\
        &M_{\mathrm{in}}\bar{U}_g,
        M_{\mathrm{in}}\bar{U}_{\ell},
        M_{\mathrm{wall}},
        M_{\mathrm{top}}
    ].
\end{split}
\label{eq:input_channels}
\end{equation}
The coordinate channels identify spatial position, the uniform-velocity channels specify the operating condition, the inlet-masked channels indicate where these conditions apply, and the wall and top masks identify the domain boundaries. Together, the four history frames and eight conditioning channels form the twelve-channel input reported in Table~\ref{tab:configuration}. For the direct phase-fraction models, we replace the SDF history with $\alpha$ while retaining the same conditioning channels.

We lift this input to a latent representation using a pointwise convolution and evolve it through Fourier Neural Operator (FNO) blocks~\cite{li2020fourier}:
\begin{equation}
    \bm{v}^{(q+1)}
    =
    \sigma
    \left[
        W_q\bm{v}^{(q)}
        +
        \mathcal{F}^{-1}
        \left(
            \mathcal{R}_q
            \odot
            \mathcal{F}\left(\bm{v}^{(q)}\right)
        \right)
    \right],
    \label{eq:fno_layer}
\end{equation}
where $\mathcal{R}_q$ contains the learned complex weights associated with the retained Fourier modes, $W_q$ is a pointwise convolution, and $\sigma$ is the GELU activation. We order the retained modes as streamwise and transverse, $(N_{k_y},N_{k_x})$, and use $(16,8)$ in the baseline model.

We compare three architectures:
\begin{align}
    \FNOsdf &:~
    \phi_{n-K+1:n}
    \longrightarrow
    \widehat{\phi}_{n+1},
    \\
    \FNOalpha &:~
    \alpha_{n-K+1:n}
    \longrightarrow
    \widehat{\alpha}_{n+1},
    \\
    \UNetalpha &:~
    \alpha_{n-K+1:n}
    \longrightarrow
    \widehat{\alpha}_{n+1}.
\end{align}
The $\FNOalpha$ model retains the conditioning and autoregressive structure of $\FNOsdf$ but predicts the phase fraction directly, whereas $\UNetalpha$ follows the implementation of Ronneberger et al.~\cite{ronneberger2015u}. We keep the dataset, history length, rollout protocol, conditioning channels, optimiser, and evaluation horizons fixed wherever the architectures permit. Parameter counts, on the other hand, are not matched across architectures, and the comparison should be read with that in mind.
To compare the SDF prediction with the phase fraction, we reconstruct a differentiable phase proxy using a logistic function:
\begin{equation}
    \mathcal{H}_{\varepsilon}(\phi)
    =
    \frac{1}{1+\exp(\phi/\varepsilon)},
    \qquad
    \varepsilon
    =
    1.5\Delta x_{\mathrm{median}},
    \label{eq:smooth_heaviside}
\end{equation}
where we adopt the convention $\phi<0$ in the liquid phase. We scale this proxy using the liquid inventory in the latest observed frame:
\begin{equation}
    A_n
    =
    \frac{
        \sum_{\x}\alpha_n(\x)
    }{
        \sum_{\x}\mathcal{H}_{\varepsilon}\!\left(\phi_n(\x)\right)
        +\reg
    },
    \qquad
    \alpha_n^p(\x)
    =
    A_n\mathcal{H}_{\varepsilon}\!\left(\phi_n(\x)\right).
    \label{eq:phase_proxy}
\end{equation}
During evaluation, ground-truth and predicted SDFs share the same $A_n$, so that phase errors are evaluated consistently on a proxy-versus-proxy basis. We stress that $A_n$ is computed from the ground-truth phase fraction of the last observed frame, so the reconstructed proxy is anchored to the true liquid inventory of the initial condition. The model is not standalone in this respect: the anchoring is available because the history itself comes from CFD. It also means that $\epsm$ measures drift away from a correctly initialised inventory rather than absolute inventory accuracy from an arbitrary start. In the design sweep of Section~\ref{sec:design_application}, where no CFD field exists at the swept point, $A_n$ is instead computed from the interpolated history, and the anchoring is only as good as that interpolation.

We train the baseline signed-distance model using an SDF reconstruction loss, an inventory penalty, and an auxiliary phase-proxy loss:
\begin{equation}
    \mathcal{L}_{\phi}
    =
    \frac{1}{R}
    \sum_{r=1}^{R}
    \left\|
        \widehat{\bar{\phi}}_{n+r}
        -
        \bar{\phi}_{n+r}
    \right\|_2^2,
    \label{eq:sdf_loss}
\end{equation}
\begin{equation}
    \mathcal{L}_m
    =
    \frac{1}{R}
    \sum_{r=1}^{R}
    \left[
        \frac{
            \widehat{M}_{n+r}^{\,p}-M_n
        }{
            |M_n|+\reg
        }
    \right]^2,
    \qquad
    M_n=\sum_{\x}\alpha_n(\x),
    \qquad
    \widehat{M}_{n+r}^{\,p}
    =
    \sum_{\x}\widehat{\alpha}_{n+r}^{\,p}(\x),
    \label{eq:inventory_loss}
\end{equation}
and
\begin{equation}
    \mathcal{L}_{\alpha}
    =
    \frac{1}{R}
    \sum_{r=1}^{R}
    \left\|
        \widehat{\alpha}_{n+r}^{\,p}
        -
        \alpha_{n+r}
    \right\|_2^2.
    \label{eq:phase_loss}
\end{equation}
Equation~\eqref{eq:inventory_loss} penalises departures from the inventory of the last observed frame, so it asks the predicted liquid content to stay constant in a domain that is continuously fed at the inlet and open at the outlet. Over the short rollouts used here the inventory changes little. Hence, the penalty acts as a stabilising anchor on an integrated quantity rather than as a statement of the correct open-domain balance. The target-increment term introduced below replaces it with the physically appropriate one.
Combining these terms, we retrieve the weighted total loss as:
\begin{equation}
    \mathcal{L}
    =
    \lambda_{\phi}\mathcal{L}_{\phi}
    +
    \lambda_m\mathcal{L}_m
    +
    \lambda_{\alpha}\mathcal{L}_{\alpha},
    \qquad
    \lambda_{\phi}=1,
    \quad
    \lambda_m=1,
    \quad
    \lambda_{\alpha}=0.05.
    \label{eq:total_loss}
\end{equation}
\paragraph{Physics-informed Neural Operator}
We then introduce a physics-regularised extension, $\FNOsdfPI$, which retains the same architecture and replaces the constant-inventory penalty with a target-increment balance:
\begin{equation}
    \mathcal{L}_{\Delta m}
    =
    \frac{1}{R}
    \sum_{r=1}^{R}
    \left[
        \frac{
            \Delta\widehat{M}_{n+r}^{\,p}
            -
            \Delta M_{n+r}^{\mathrm{GT}}
        }{
            |M_n^{\mathrm{GT}}|+\reg
        }
    \right]^2,
    \label{eq:target_increment_loss}
\end{equation}
where
\begin{equation}
    \Delta\widehat{M}_{n+r}^{\,p}
    =
    \widehat{M}_{n+r}^{\,p}
    -
    \widehat{M}_{n+r-1}^{\,p},
    \qquad
    \Delta M_{n+r}^{\mathrm{GT}}
    =
    M_{n+r}^{\mathrm{GT}}
    -
    M_{n+r-1}^{\mathrm{GT}}.
\end{equation}
This term accounts for changes in the open-domain inventory, although it remains supervised by the CFD target rather than evaluated using boundary fluxes.

We also regularise the metric property of the predicted SDF. Since an exact signed-distance field satisfies $|\nabla\phi|=1$, we penalise deviations from this condition within a narrow band around the interface \cite{gropp2020implicit,yariv2021volume}:
\begin{equation}
    \mathcal{L}_{\mathrm{Eik}}
    =
    \frac{1}{R}
    \sum_{r=1}^{R}
    \frac{
        \sum_{\x}
        w_{\Gamma,n+r}(\x)
        \left(
            |\nabla\widehat{\phi}_{n+r}(\x)|-1
        \right)^2
    }{
        \sum_{\x}w_{\Gamma,n+r}(\x)+\reg
    },
    \label{eq:eikonal_loss}
\end{equation}
where
\begin{equation}
    w_{\Gamma}(\x)
    =
    \exp\left(
        -\frac{|\widehat{\phi}(\x)|}{\beta_{\Gamma}}
    \right),
    \qquad
    \beta_{\Gamma}=4\varepsilon.
\end{equation}
We evaluate this term using the dimensional, denormalised SDF and calculate spatial derivatives using centred second-order differences. Finally, we discourage the reconstructed phase proxy from leaving its admissible interval:
\begin{equation}
    \mathcal{L}_{\mathrm{bnd}}
    =
    \frac{1}{R}
    \sum_{r=1}^{R}
    \left\langle
        \operatorname{ReLU}
        \left(
            -\widehat{\alpha}_{n+r}^{\,p}
        \right)^2
        +
        \operatorname{ReLU}
        \left(
            \widehat{\alpha}_{n+r}^{\,p}-1
        \right)^2
    \right\rangle_{\Omega}.
    \label{eq:boundedness_loss}
\end{equation}
Because the logistic reconstruction of Equation~\eqref{eq:smooth_heaviside} already returns an admissible field and the scale factor $A_n$ is close to unity, this term evaluated to zero throughout training. It therefore acts as a guarantee that admissibility is not violated rather than as a term that shapes the solution, and it should be read that way in the results below.
The resulting objective is
\begin{equation}
    \mathcal{L}_{\mathrm{PI}}
    =
    \lambda_{\phi}\mathcal{L}_{\phi}
    +
    \lambda_{\Delta m}\mathcal{L}_{\Delta m}
    +
    \lambda_{\alpha}\mathcal{L}_{\alpha}
    +
    \lambda_{\mathrm{Eik}}\mathcal{L}_{\mathrm{Eik}}
    +
    \lambda_{\mathrm{bnd}}\mathcal{L}_{\mathrm{bnd}},
    \label{eq:physics_informed_total_loss}
\end{equation}
with
\begin{equation}
    \lambda_{\phi}=1,
    \qquad
    \lambda_{\Delta m}=1,
    \qquad
    \lambda_{\alpha}=0.05,
    \qquad
    \lambda_{\mathrm{Eik}}=0.05,
    \qquad
    \lambda_{\mathrm{bnd}}=0.01.
\end{equation}

\paragraph{Training protocol} We train all models with AdamW using autoregressive rollouts, with the settings collected in Table~\ref{tab:configuration}. After receiving the ground-truth history, each model appends its prediction to the input sequence and uses it to predict the next field, thereby learning under its own accumulated errors.

\paragraph{Evaluation metrics}
We evaluate the models using field accuracy, phase accuracy, liquid-inventory conservation, and interface overlap. For the signed-distance models, we calculate the dimensional SDF error as
\begin{equation}
    \rmsephi(\Dn)
    =
    \sqrt{
        \frac{1}{N_xN_y}
        \sum_{\x}
        \left[
            \widehat{\phi}_{n+\Dn}(\x)
            -
            \phi_{n+\Dn}^{\mathrm{GT}}(\x)
        \right]^2
    },
    \label{eq:rmse_phi}
\end{equation}
and calculate $\rmsealphap$ between the predicted and ground-truth phase proxies obtained from Equation~\eqref{eq:phase_proxy}. For the direct phase-fraction models, we instead compute $\rmsealpha$ relative to the raw CFD phase fraction. The proxy is smoother than the raw phase fraction on both sides of the signed-distance comparison, so the two error families are not measured in the same target space. Consequently, the comparison combines differences in both architecture and field representation and should not be interpreted as a strict common-target ranking.

We quantify the relative liquid-inventory error as
\begin{equation}
    \epsm(\Dn)
    =
    \frac{
        \left|
            \sum_{\x}\widehat{\alpha}_{n+\Dn}^{\,\star}(\x)
            -
            \sum_{\x}\alpha_{n+\Dn}^{\star,\mathrm{GT}}(\x)
        \right|
    }{
        \sum_{\x}\alpha_{n+\Dn}^{\star,\mathrm{GT}}(\x)
        +
        \reg
    },
    \label{eq:mass_metric}
\end{equation}
where $\alpha^\star=\alpha^p$ for the signed-distance models and $\alpha^\star=\alpha$ for the direct phase-fraction models. Unlike the baseline inventory loss in Equation~\eqref{eq:inventory_loss}, this metric uses the ground-truth inventory at the evaluation time. We finally calculate the intersection over union of the liquid regions \cite{rahman2016optimizing}:
\begin{equation}
    \mathrm{IoU}(\Dn)
    =
    \frac{
        |\Omega_{\mathrm{pred}}(\Dn)\cap\Omega_{\mathrm{GT}}(\Dn)|
    }{
        |\Omega_{\mathrm{pred}}(\Dn)\cup\Omega_{\mathrm{GT}}(\Dn)|
    }.
    \label{eq:iou}
\end{equation}

\paragraph{Interfacial area and design observables}
\label{sec:interfacial_area}

We now introduce the observables proposed for applying the model to optimal nozzle characteristics. The first parameter we extract is the specific interfacial area, which determines the available rate of interphase heat and mass transfer. The signed-distance representation gives this directly: the interface is the zero level set of $\phi$, so its extent follows from a contour extraction on a field the model already predicts, with no thresholding step. In the two-dimensional domain used here,
\begin{equation}
    a_{\mathrm{2D}}(\Dn)
    =
    \frac{
        \left|
            \left\{
                \x\in\Omega : \phi_{n+\Dn}(\x)=0
            \right\}
        \right|
    }{
        |\Omega|
    },
    \label{eq:interfacial_area}
\end{equation}
where the numerator is the total length of the zero contour and $|\Omega|$ the domain area, so that $a_{\mathrm{2D}}$ has dimensions of inverse length. Because the domain is two-dimensional, $a_{\mathrm{2D}}$ is a perimeter per unit area rather than a true interfacial area per unit volume. We could have extracted this quantity directly from the geometric reconstruction performed by OpenFOAM's \texttt{isoAdvector} at runtime, but we did not do so to limit computational cost. The other parameter of interest for our analysis, given the airblast configuration, is the kinetic power delivered by the gas stream, which scales as $\rho_g A_g \Ug^3$, so with fixed gas density and annulus area, we use a normalised proxy
\begin{equation}
    P^*_{\mathrm{gas}}
    =
    \left(
        \frac{\Ug}{\Ug^{\mathrm{ref}}}
    \right)^{\!3},
    \qquad
    \Ug^{\mathrm{ref}}=40,
    \label{eq:gas_power}
\end{equation}
and finally define the atomisation efficiency as the interfacial area generated per unit gas power,
\begin{equation}
    \eta^*
    =
    \frac{\widehat{a}_{\mathrm{2D}}}{P^*_{\mathrm{gas}}}.
    \label{eq:efficiency}
\end{equation}
Equation~\eqref{eq:gas_power} should not be interpreted as a measured compressor duty, and the efficiency map can be re-derived under a different cost assumption.
The total interfacial length is not a suitable design parameter on its own, as it does not distinguish between liquid that has been atomised and liquid that remains in an intact column. Therefore, a long, smooth, unbroken jet and a fully fragmented spray might yield comparable $a_{\mathrm{2D}}$ values yet represent entirely different atomisation states. On the other hand, the latter provides the dispersed area required for downstream transfer, suggesting, therefore, that the predicted liquid region is decomposed by topology.
We remove isolated components smaller than 4 pixels as segmentation noise and partition the remainder into connected components using eight-connectivity. Components intersecting the first two rows at the injection boundary form the intact \emph{core} $\Omega_{\mathrm{core}}$; every remaining component of at least eight pixels is a \emph{detached} structure, and their union is $\Omega_{\mathrm{det}}$. Unattached components of four to seven pixels fall below the size at which a fragment is counted and are discarded, so the decomposition
\begin{equation}
    a_{\mathrm{2D}}^{\mathrm{total}}
    =
    a_{\mathrm{2D}}^{\mathrm{core}}
    +
    a_{\mathrm{2D}}^{\mathrm{det}}
    \label{eq:area_decomposition}
\end{equation}
is exact over the two \emph{retained} partitions rather than over the full zero contour of Equation~\eqref{eq:interfacial_area}; $a_{\mathrm{2D}}^{\mathrm{total}}$ is used in this retained sense throughout the topology maps. The discarded sub-eight-pixel material is one reason the fragmentation diagnostics below are lower bounds. Detached structures truncated by the domain boundary are, by contrast, retained. Hence, the decomposition measures the interface visible within the region of interest rather than attempting to reconstruct fragments that have left it. Three further topology diagnostics follow: the detached area fraction $f_{\mathrm{det}}=A_{\mathrm{det}}/A_{\mathrm{liq}}$, the number of detached structures $N_{\mathrm{det}}$, and the normalised core penetration $L_{\mathrm{core}}/L_{\mathrm{ROI}}$, the deepest streamwise index reached by the intact core relative to the domain height.

An operating point is classified as \emph{atomising} when the liquid has visibly separated at all,
\begin{equation}
    f_{\mathrm{det}}
    \;\geq\;
    f_{\mathrm{det}}^{\mathrm{crit}}=2\times10^{-3}
    \quad\text{and}\quad
    N_{\mathrm{det}}\geq1.
    \label{eq:atomisation_onset}
\end{equation}
The locus of this condition defines the atomisation-onset boundary. The criterion is deliberately permissive: it asks only that at least one detached structure of eight pixels or more exists and that it carries at least $0.2\%$ of the liquid area. The design objective is the detached interfacial area generated per unit gas power, evaluated only on the feasible set,
\begin{equation}
    \eta^*_{\mathrm{atom}}
    =
    \frac{\widehat{a}_{\mathrm{2D}}^{\,\mathrm{det}}}{P^*_{\mathrm{gas}}}.
\end{equation}
A complementary objective credits, in addition, any core interface created beyond that of a low-gas-velocity reference at the same liquid throughput,
\begin{equation}
    a_{\mathrm{2D}}^{\mathrm{gen}}
    =
    a_{\mathrm{2D}}^{\mathrm{det}}
    +
    \max\!\left(
        a_{\mathrm{2D}}^{\mathrm{core}}
        -
        a_{\mathrm{2D}}^{\mathrm{core,ref}},\,
        0
    \right),
    \qquad
    \Ug^{\mathrm{ref}}=40,
    \label{eq:generated_area}
\end{equation}
which rewards corrugation of a column that has not yet fully broken up. The reference at $\Ug^{\mathrm{ref}}=40$ is the least fragmented condition in the envelope rather than a non-atomising one, so Equation~\eqref{eq:generated_area} measures interface created \emph{relative} to that baseline state. Results below use $\eta^*_{\mathrm{atom}}$; $a_{\mathrm{2D}}^{\mathrm{gen}}$ is reported as a supplementary field. Equation~\eqref{eq:efficiency}, built on undifferentiated total area, is retained only as the unconstrained comparator that motivates the refinement.

\section{Results}
\label{sec:results}

\begin{figure}[!htbp]
    \centering
    \begin{subfigure}[t]{0.48\linewidth}
        \centering
        \includegraphics[width=\linewidth]{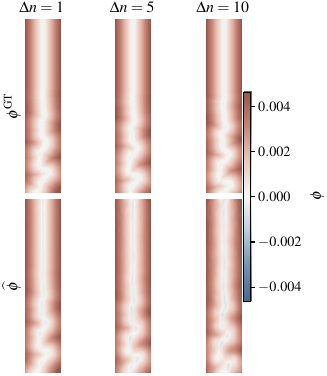}
        \caption{Signed-distance field $\phi$.}
    \end{subfigure}
    \hfill
    \begin{subfigure}[t]{0.48\linewidth}
        \centering
        \includegraphics[width=\linewidth]{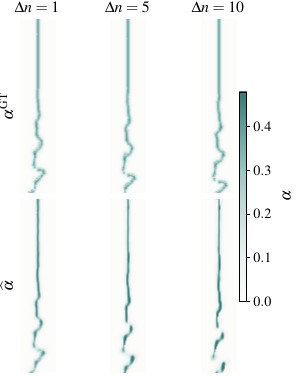}
        \caption{SDF-derived phase proxy $\alpha^p$.}
    \end{subfigure}
    \caption{
    Ground-truth and autoregressive $\FNOsdf$ predictions for $\Ug=50$, $\Ul=7$
    at $\Dn=1$, $5$, and $10$. The upper row of each subfigure is the ground
    truth and the lower row the prediction. This operating condition is part of the training set $\Dtrain$, so the panels show in-sample forecast quality.}
    \label{fig:case_50_7}
\end{figure}

\begin{figure}[!htbp]
    \centering
    \begin{subfigure}[t]{0.48\linewidth}
        \centering
        \includegraphics[width=\linewidth]{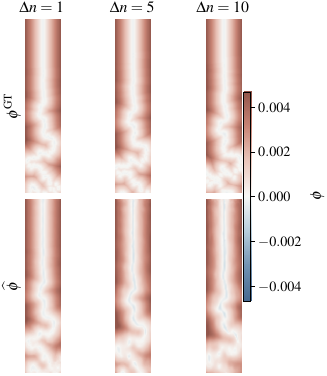}
        \caption{Signed-distance field $\phi$.}
    \end{subfigure}
    \hfill
    \begin{subfigure}[t]{0.48\linewidth}
        \centering
        \includegraphics[width=\linewidth]{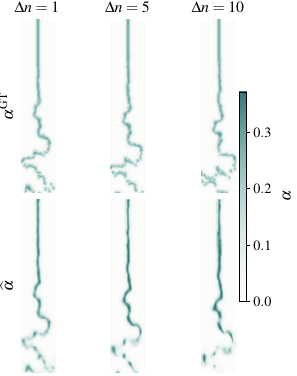}
        \caption{SDF-derived phase proxy $\alpha^p$.}
    \end{subfigure}
    \caption{
    Ground-truth and autoregressive $\FNOsdf$ predictions for $\Ug=70$, $\Ul=9$
    at $\Dn=1$, $5$, and $10$. Stronger fragmentation makes isolated structures
    harder to retain over the rollout. As in Figure~\ref{fig:case_50_7}, this
    operating condition is part of the training set $\Dtrain$.}
    \label{fig:case_70_9}
\end{figure}

Figure~\ref{fig:case_50_7} shows a comparison of the predicted spray morphology and the ground truth for two cases that are part of the training set. The model captures the volume fraction well when the liquid column is well-defined and moderately deformed. The main differences are, in fact, small detached structures and local curvature at longer horizons. As the velocities of liquid and gas increase, as depicted in Figure~\ref{fig:case_70_9}, the flow is more strongly deformed and fragmented in the upper domain, thus more challenging for the model to capture. However, the model still captures the liquid-core trajectory and the location of the principal structures.

\begin{figure}[!htbp]
    \centering
    \begin{subfigure}[t]{0.48\linewidth}
        \centering
        \includegraphics[width=\linewidth]{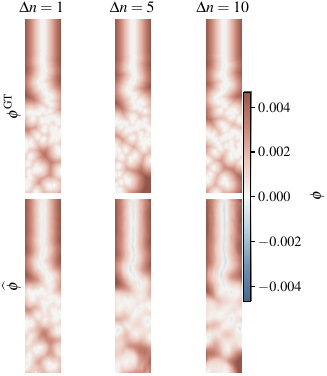}
        \caption{Signed-distance field $\phi$.}
    \end{subfigure}
    \hfill
    \begin{subfigure}[t]{0.48\linewidth}
        \centering
        \includegraphics[width=\linewidth]{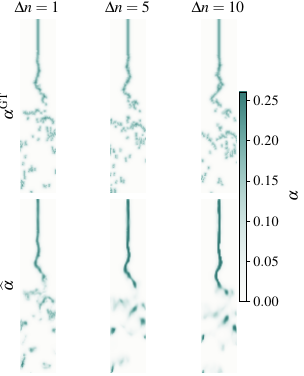}
        \caption{SDF-derived phase proxy $\alpha^p$.}
    \end{subfigure}
    \caption{
    High-velocity extrapolation with $\FNOsdf$ for $\Ug=90$, $\Ul=8$ at
    $\Dn=1$, $5$, and $10$. Upper rows are ground truth, lower rows are
    predictions. This condition belongs to $\Dtest$ and was excluded from
    training.}
    \label{fig:case_90_8}
\end{figure}

\FloatBarrier
If we look at the case involving higher velocities and extrapolated to an unexplored region of the parameter space, the model starts to lose fidelity. Nevertheless, Figure~\ref{fig:case_90_8} shows that the predictions made by the $\FNOsdf$ model are still acceptable, particularly if one looks at the SDF instead of the retrieved $\alpha$ proxy.

\FloatBarrier

\subsection{Architecture and representation comparison}
\label{sec:architecture_comparison}

\begin{figure}[!htbp]
    \centering
    \includegraphics[width=\linewidth]{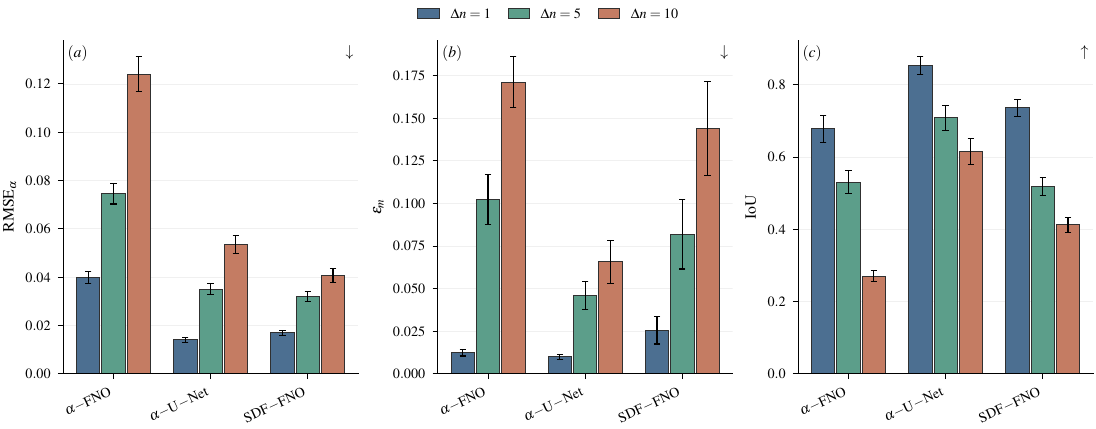}
    \caption{
    Comparison of direct-phase Fourier forecasting ($\FNOalpha$), direct-phase
    U-Net forecasting ($\UNetalpha$), and signed-distance Fourier forecasting
    ($\FNOsdf$) at $\Dn=1$, $5$, and $10$: (a) phase-field error, (b) relative
    liquid-inventory error $\epsm$, (c) interface overlap $\mathrm{IoU}$.
    Phase-field error is $\rmsealpha$ against raw $\alpha$ for the direct-phase
    models and $\rmsealphap$ proxy-versus-proxy for $\FNOsdf$; the two occupy
    the same axis but are measured in different target spaces. Arrows in the panel corners indicate whether lower
    ($\downarrow$) or higher ($\uparrow$) is better.}
    \label{fig:architecture_comparison}
\end{figure}

Figure~\ref{fig:architecture_comparison} compares the direct-phase FNO ($\FNOalpha$), direct-phase U-Net ($\UNetalpha$), and signed-distance FNO ($\FNOsdf$). Panels (a) and (b) report phase-field error and relative inventory error; both are threshold-independent, but we scored the signed-distance model proxy-versus-proxy while the direct-phase models are scored against the raw CFD field. Panel (c) reports interface overlap (IoU).
We highlight two things: first, on a fixed grid (the case in object), the direct-phase U-Net is the strongest pixel-level predictor, performing best across all horizons.
The second is that replacing the discontinuous phase fraction with the signed-distance field brings the same operator down to $\rmsealphap=0.0171$, $0.0320$, and $0.0408$ over those horizons, roughly a threefold reduction at $\Dn=10$ relative to the direct-phase FNO. The proxy is a smoothed field on both sides of that ratio, so the factor of three quantifies the benefit of predicting a geometric field and reconstructing the phase from it. With that qualification, spectral operator learning becomes markedly more effective when the interface is encoded geometrically rather than as a discontinuous indicator.

\FloatBarrier

\subsection{Ablation of the training objective}
\label{sec:ablation}

To better identify the role of each term in the quality of the prediction, we performed an ablation study. In particular, we evaluated the effect of the two auxiliary loss terms and the retained spectral bandwidth. Relative to the baseline objective of Equation~\eqref{eq:total_loss}, $\mathcal{L}_{\phi}+\mathcal{L}_m+0.05\mathcal{L}_{\alpha}$, four variants are trained: $\lambda_m=0$, which removes the liquid-inventory penalty; $\lambda_{\alpha}=0$, which removes the auxiliary phase-proxy loss; $\mathcal{L}=\mathcal{L}_{\phi}$, which retains signed-distance supervision only; and $N_k\downarrow$, which reduces the number of retained Fourier modes from the baseline $(16,8)$.

\begin{figure}[!htbp]
    \centering
    \includegraphics[width=\linewidth]{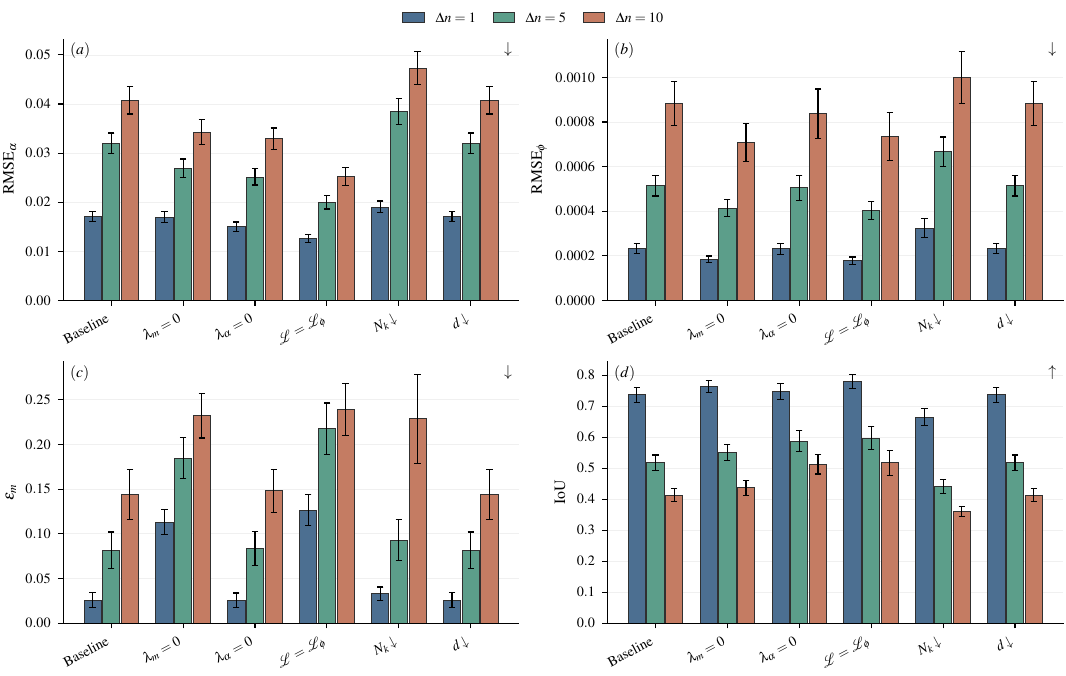}
    \caption{
    Case-averaged ablation metrics for the $\FNOsdf$ variants at $\Dn=1$, $5$,
    and $10$: (a) signed-distance error $\rmsephi$, (b) relative liquid-inventory
    error $\epsm$, (c) interface overlap $\mathrm{IoU}$. Arrows indicate whether
    lower ($\downarrow$) or higher ($\uparrow$) is better. All variants share an
    identical per-case overlap threshold, so every panel is a like-for-like
    comparison. Error bars are the standard error across the $60$ evaluated
    cases.}
    \label{fig:ablation_mosaic}
\end{figure}

\emph{Signed-distance and phase accuracy.} $\rmsephi$ rises monotonically with horizon for every model, reflecting accumulated rollout error (Figure~\ref{fig:ablation_mosaic}a). Reducing the retained modes yields the largest error across all horizons, reaching $9.99\times10^{-4}$ at $\Dn=10$, compared with a baseline of $8.81\times10^{-4}$. The $\mathcal{L}=\mathcal{L}_{\phi}$ and $\lambda_m=0$ configurations give the smallest errors, $1.78\times10^{-4}$ and $1.84\times10^{-4}$ at $\Dn=1$ against a baseline of $2.33\times10^{-4}$ (standard errors of order $1.5\times10^{-5}$). Removing the auxiliary phase loss leaves $\rmsephi$ essentially unchanged at $\Dn=1$ and slightly improved at longer horizons.

\emph{Liquid inventory.} The inventory error gives the clearest separation in the ablation (Figure~\ref{fig:ablation_mosaic}b). The baseline and $\lambda_{\alpha}=0$ configurations, both retaining $\mathcal{L}_m$, reach $\epsm=0.026$ at $\Dn=1$. Removing the penalty raises this to $0.113$ for $\lambda_m=0$ and $0.127$ for $\mathcal{L}=\mathcal{L}_{\phi}$, which is roughly a fivefold increase at that horizon. The gap narrows as the rollout lengthens, and by $\Dn=10$ the unregularised configurations reach roughly $23\%$ compared with about $14\%$ for the baseline, a factor of roughly $1.6$.

\emph{Interface overlap.} The overlap results track the signed-distance errors (Figure~\ref{fig:ablation_mosaic}c). IoU decreases with horizon for every configuration, and the reduced-mode model gives the lowest overlap throughout, falling to $0.361\pm0.016$ at $\Dn=10$ against a baseline of $0.413\pm0.021$; truncating the spectral representation clearly impairs retention of thin ligaments. Among the remaining four configurations the differences are small relative to the case-to-case scatter: at $\Dn=1$ the values span $0.736\pm0.024$ (baseline) to $0.779\pm0.023$ ($\mathcal{L}=\mathcal{L}_{\phi}$), and at $\Dn=10$ from $0.413\pm0.021$ (baseline) to $0.517\pm0.040$ ($\mathcal{L}=\mathcal{L}_{\phi}$). The configurations without inventory regularisation sit at or above the baseline, but their intervals overlap substantially, and we claim no strict ordering.

The ablation shows something extremely relevant: the geometrically most accurate model is not the most physically conservative one. The $\mathcal{L}=\mathcal{L}_{\phi}$ and $\lambda_m=0$ configurations improve $\rmsephi$ and are at least as good on overlap. However, their inventory errors are several times larger than the baseline at short horizons. The model we propose is therefore a deliberate compromise between local interface fidelity and global inventory consistency. Because the inventory effect is large relative to its uncertainty, whereas the overlap differences are not, the inventory effect is the more robust of the two. We note that the $N_k\downarrow$ variant establishes only that reducing bandwidth below $(16,8)$ degrades the prediction; it says nothing about what would be gained by increasing it, which we did not test.

\FloatBarrier

\subsection{Physics-informed extension}
\label{sec:physics_informed}

We then sought to determine whether the inverse relationship between local interface fidelity and inventory conservation exposed by the ablation could be improved by the physics-informed extension of Section~\ref{sec:models}, which replaces the constant-inventory penalty with the target-increment balance of Equation~\eqref{eq:target_increment_loss} and adds the Eikonal and boundedness terms of Equations~\eqref{eq:eikonal_loss} and~\eqref{eq:boundedness_loss}. The extension was trained on the same $50$ conditions as the baseline, with all $\Ug=90$ cases excluded. It has $794{,}561$ trainable parameters and the same twelve input channels as the baseline, so any difference in behaviour is attributable to the objective rather than to capacity or conditioning.

\emph{Training behaviour.} Table~\ref{tab:physics_training_history} summarises the first and final epochs. The total objective falls from $1.53\times10^{-2}$ to $3.63\times10^{-3}$, about a factor of four, and the monitored objective from $8.46\times10^{-3}$ to $3.78\times10^{-3}$, with the target-increment term dropping by more than two orders of magnitude and the Eikonal term decreasing throughout. The boundedness penalty remains zero, so the reconstructed proxy stays within $[0,1]$ across the monitored batches, confirming that this term guarantees admissibility rather than shaping the solution. These curves establish that the combined objective optimises stably, but because the monitoring windows overlap the training set, they are a convergence diagnostic rather than a generalisation result.

\begin{table}[!htbp]
    \centering
    \caption{Training and monitoring terms for the physics-informed run. The monitoring subset overlaps the training set and serves only as a convergence diagnostic.}
    \label{tab:physics_training_history}
    \begin{tabular}{lrrrr}
        \toprule
        & \multicolumn{2}{c}{Epoch 1} & \multicolumn{2}{c}{Epoch 50} \\
        \cmidrule(lr){2-3}\cmidrule(lr){4-5}
        Term & Train & Monitor & Train & Monitor \\
        \midrule
        $\mathcal{L}_{\mathrm{PI}}$
        & $1.5326\times10^{-2}$ & $8.4619\times10^{-3}$
        & $3.6312\times10^{-3}$ & $3.7778\times10^{-3}$ \\
        $\mathcal{L}_{\phi}$
        & $6.9999\times10^{-3}$ & $4.8857\times10^{-3}$
        & $1.5323\times10^{-3}$ & $1.6829\times10^{-3}$ \\
        $\mathcal{L}_{\Delta m}$
        & $5.7321\times10^{-3}$ & $1.0257\times10^{-3}$
        & $2.8730\times10^{-5}$ & $2.8201\times10^{-5}$ \\
        $\mathcal{L}_{\alpha}$
        & $1.2500\times10^{-2}$ & $1.2027\times10^{-2}$
        & $1.1651\times10^{-2}$ & $1.1323\times10^{-2}$ \\
        $\mathcal{L}_{\mathrm{Eik}}$
        & $3.9372\times10^{-2}$ & $3.8983\times10^{-2}$
        & $2.9754\times10^{-2}$ & $3.0011\times10^{-2}$ \\
        \bottomrule
    \end{tabular}
\end{table}

\begin{figure}[!htbp]
    \centering
    \includegraphics[width=\linewidth]{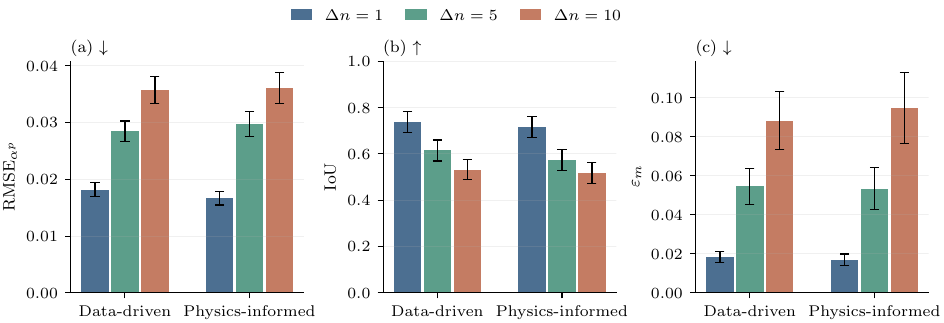}
    \caption{
    Case-wide comparison between the data-driven $\FNOsdf$ and physics-informed
    $\FNOsdfPI$ signed-distance operators, averaged over the $60$ evaluated cases,
    at $\Dn=1$, $5$, and $10$: (a) phase-proxy error $\rmsealphap$,
    (b) interface overlap $\mathrm{IoU}$, (c) relative liquid-inventory error
    $\epsm$. Arrows indicate whether lower ($\downarrow$) or higher ($\uparrow$)
    is better. Of the $60$ cases, $50$ are the training conditions themselves and
    $10$ are the held-out $\Ug=90$ strip. The two operators are separated by less
    than the case-to-case standard error on every metric and at every horizon.}
    \label{fig:physics_comparison_casewide}
\end{figure}

\emph{Predictive comparison.} Whether that optimisation translates into better forecasting is a separate question, and it is the one that matters for the design application. We therefore compare $\FNOsdf$ and $\FNOsdfPI$ directly at the predictive level over the $60$ evaluated cases, scoring both on identical initial histories and rollout horizons with $\rmsealphap$, $\epsm$, and $\mathrm{IoU}$ at $\Dn\in\{1,5,10\}$. Figure~\ref{fig:physics_comparison_casewide} shows the outcome: the two operators track one another on all three metrics and at every horizon, and the separation between them is in each case smaller than the standard error across cases. This dataset therefore resolves no predictive difference between the data-driven and the physics-informed objective, despite the latter reducing its own training residuals substantially. The additional geometric and boundedness regularity comes at no predictive cost, but it also buys no predictive gain, and we carry the baseline $\FNOsdf$ forward into the design application of Section~\ref{sec:design_application} on that basis.

\FloatBarrier

\section{Application: rapid exploration of injector operating conditions}
\label{sec:design_application}

While the previous sections describe the surrogate model and various alternatives, the aim of the current section is to demonstrate a potential application of the developed architecture. We use the model to identify the optimal velocities at which operating the spray is economically favourable. In practice we look for maximising the available surface area while keeping the power required to move the gas at a minimum.
This application is particularly relevant for this paper since exploring such a space would be undoable with a full CFD simulation, even if limited to two dimensions.

The following calculation shows how convenient substituting CFD with a data-driven surrogate is. The space explored in this application comprises $3000$ operating points and completes in $173$~s, or $57.7$~ms per design point, and the full $60$-case validation rollout costs $5.1$~s in total. The reference \texttt{isoAdvector} simulations average $32$~minutes per transient case, so covering the same $3000$ points with CFD would require about $5.8\times10^{6}$~s of solver time, roughly $67$~days, against under three minutes of inference. The marginal cost of interrogating one additional operating point is therefore reduced by a factor of approximately $3.3\times10^{4}$, roughly four orders of magnitude.

\begin{figure}[!htbp]
    \centering
    \includegraphics[width=\linewidth]{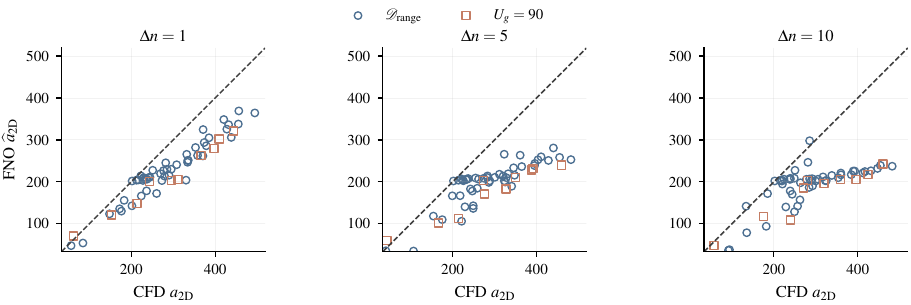}
    \caption{
    Predicted against reference interfacial area $a_{\mathrm{2D}}$ for every evaluated case at $\Dn=1$, $5$, and $10$. Circles are the in-sample training
    conditions $\Drange$ and squares the $\Ug=90$ held-out conditions $\Dtest$; the dashed line is $1{:}1$. The surrogate under-predicts interfacial area at every condition and compresses its dynamic range with increasing horizon, but preserves the ordering of operating points ($\rho=0.93$, $0.80$, $0.73$ at $\Dn=1$, $5$, $10$), which is the property the design screening of Figures~\ref{fig:design_maps} and~\ref{fig:pareto} relies on.}
    \label{fig:area_validation}
\end{figure}

\emph{Validation of the design observable.} A design map is only as trustworthy as the quantity it maps, so we first compare predicted and reference interfacial area case by case (Figure~\ref{fig:area_validation}). It is evident that the surrogate systematically \emph{under-predicts} $a_{\mathrm{2D}}$: every evaluated case falls below the $1{:}1$ line, with a mean ratio $\widehat{a}_{\mathrm{2D}}/a_{\mathrm{2D}}$ of about $0.74$ at $\Dn=1$, falling to $0.65$ at $\Dn=5$ and $0.63$ at $\Dn=10$. The regression slope contracts correspondingly, from about $0.67$ to $0.39$, so the predicted dynamic range is compressed as well as shifted. This is the direct signature of the interface smoothing documented in Section~\ref{sec:results}: interfacial area is dominated by small ligaments and detached fragments, which are exactly the structures the spectral representation and the mean-square objective suppress first.

What survives the bias is the ordering. The rank correlation between predicted and reference area is $\rho=0.93$ at $\Dn=1$ ($R^2=0.90$), $\rho=0.80$ at $\Dn=5$, and $\rho=0.73$ at $\Dn=10$. The surrogate is thus a biased but largely rank-preserving estimator of interfacial area, which is the property a screening tool requires: it can order candidate operating points even where it cannot quantify them. Since $50$ of the $60$ cases are training conditions, this rank fidelity is established predominantly in sample; the $\Dtest$ cases at $\Ug=90$ are marked separately in Figure~\ref{fig:area_validation}. Each point is a single forecast window, so the scatter reflects the snapshot-to-snapshot fluctuations in interfacial area in a statistically stationary spray and is not a pure model error.

\begin{figure}[!htbp]
    \centering
    \includegraphics[width=\linewidth]{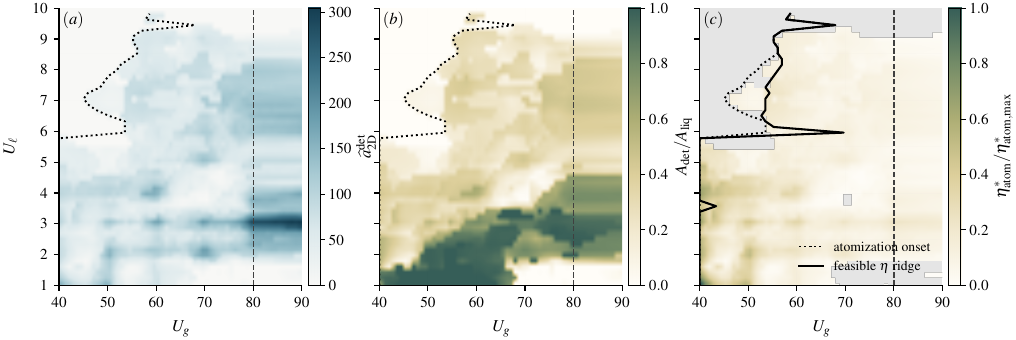}
    \caption{
    Atomisation-aware design maps over the $(\Ug,\Ul)$ operating envelope, evaluated at $\Dn=10$: (a) detached interfacial density $\widehat{a}_{\mathrm{2D}}^{\,\mathrm{det}}$, (b) detached area fraction $A_{\mathrm{det}}/A_{\mathrm{liq}}$, (c) constrained atomisation efficiency
    $\eta^*_{\mathrm{atom}}$, normalised by its maximum. The dotted contour is the atomisation-onset boundary of Equation~\eqref{eq:atomisation_onset}; shading in panel (c) marks the infeasible region where the liquid does not break up, and the solid line is the feasible efficiency ridge. The dashed line at $\Ug=80$ separates the trained envelope from the extrapolation region; to its right, the field history is clamped to the $\Ug=80$ condition, so the variation shown there is the response of the conditioning channels alone.}
    \label{fig:design_maps}
\end{figure}

\emph{Construction of the swept histories.} The operator requires a $K$-frame field history in addition to the conditioning pair, and no such history exists at operating points that were never simulated. Histories at swept points are therefore constructed by bilinear interpolation of developed CFD histories from the surrounding four training conditions, with the requested $(\Ug,\Ul)$ retained in the conditioning channels; beyond the training envelope the history is clamped to the nearest edge, so that the $\Ug>80$ strip uses the $\Ug=80$ history throughout and varies only through conditioning. Predicted interfacial area is biased low between nodes relative to at them, which contributes visible structure aligned with the training grid in Figure~\ref{fig:design_maps}. The maps should therefore be read as a coarse response surface over the envelope rather than as a resolved field, and the fine-scale texture between training conditions should not be interpreted physically.

\begin{figure}[!htbp]
    \centering
    \includegraphics[width=\linewidth]{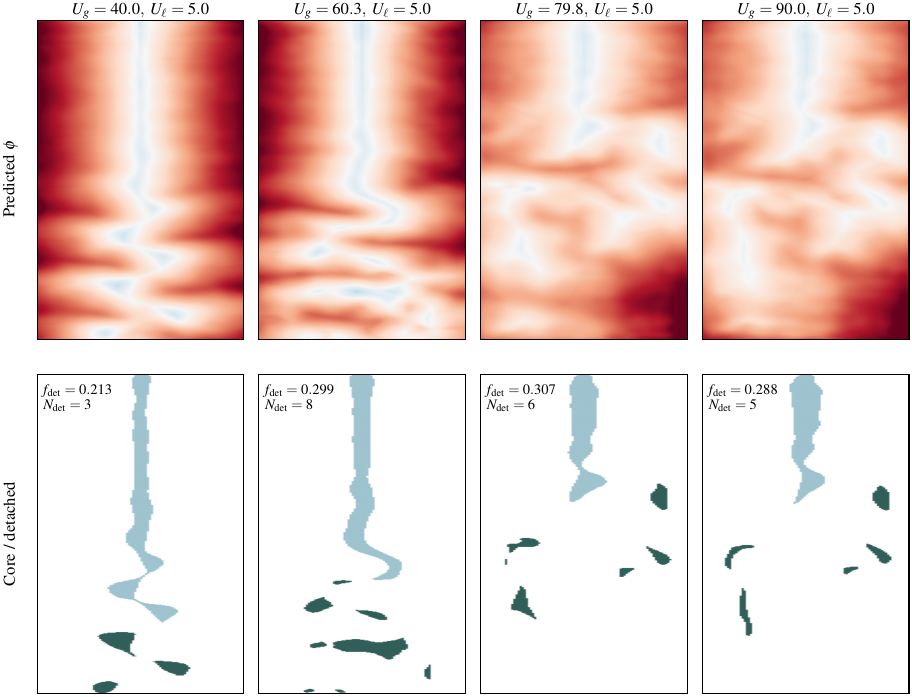}
    \caption{
    Topology decomposition of the predicted field along the constant-$\Ul=5$
    line at $\Dn=10$. Upper row: predicted signed-distance field $\widehat{\phi}$.
    Lower row: the corresponding liquid region partitioned into the intact core
    attached to the injection boundary (light) and detached structures (dark),
    annotated with the detached area fraction $f_{\mathrm{det}}$ and the number
    of detached structures $N_{\mathrm{det}}$. The intact column shortens as
    $\Ug$ increases, and the detached area fraction rises across the trained
    range, which is the transition that the atomisation-onset criterion of
    Equation~\eqref{eq:atomisation_onset} is constructed to detect; the count
    $N_{\mathrm{det}}$ is not monotone because fragments consolidate and leave
    the region of interest at the highest gas velocities. All four columns lie
    far above the onset threshold, so none of them probes the boundary itself.}
    \label{fig:topology_diagnostics}
\end{figure}

\begin{figure}[!htbp]
    \centering
    \includegraphics[width=\linewidth]{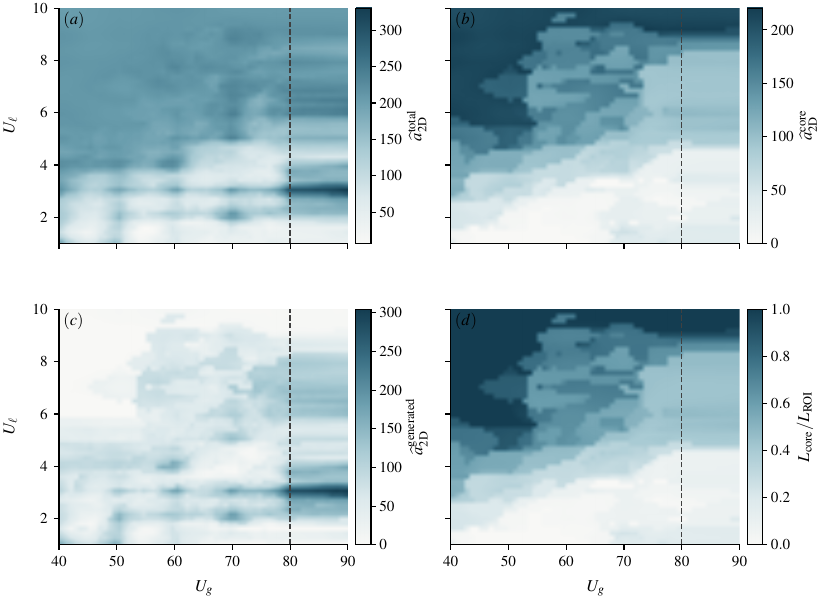}
    \caption{
    Supplementary topology maps over the operating envelope at $\Dn=10$:
    (a) retained total interfacial density
    $\widehat{a}_{\mathrm{2D}}^{\,\mathrm{total}}$ of
    Equation~\eqref{eq:area_decomposition},
    (b) core contribution $\widehat{a}_{\mathrm{2D}}^{\,\mathrm{core}}$,
    (c) generated interfacial density $\widehat{a}_{\mathrm{2D}}^{\,\mathrm{gen}}$
    of Equation~\eqref{eq:generated_area}, which adds to the detached interface
    any core interface in excess of the $\Ug=40$ reference at the same $\Ul$,
    (d) normalised core penetration $L_{\mathrm{core}}/L_{\mathrm{ROI}}$. The comparative flatness of the total in panel (a) conceals the systematic transfer of interfacial length from the intact core to detached fragments visible in panels (b) and (d).}
    \label{fig:topology_supplementary}
\end{figure}

\emph{Interface topology across the envelope.} Before mapping an objective we separate the intact core from detached fragments, because the two respond to the operating conditions in opposite ways. Figure~\ref{fig:topology_diagnostics} illustrates the decomposition along the constant-$\Ul=5$ line at four gas velocities. At $\Ug=40$ the predicted liquid region is dominated by a long column that traverses most of the domain and carries close to four-fifths of the liquid area, with only three detached structures and $f_{\mathrm{det}}=0.213$; the column is the least fragmented state in the envelope rather than a strictly unbroken one, which is why it serves as the reference in Equation~\eqref{eq:generated_area}. As the gas velocity rises the core retracts and the field fragments: at $\Ug=60.3$ the detached fraction reaches $f_{\mathrm{det}}=0.299$ with $N_{\mathrm{det}}=8$ structures, and by $\Ug=79.8$ the core terminates in the upper domain with $f_{\mathrm{det}}=0.307$ and $N_{\mathrm{det}}=6$. The extrapolated $\Ug=90$ column shows $f_{\mathrm{det}}=0.288$ and $N_{\mathrm{det}}=5$. The detached \emph{area fraction} therefore rises across the trained range and then falls back slightly at the extrapolation edge, where the history is, in any case, clamped to $\Ug=80$. In contrast, the \emph{count} is not monotone ($3$, $8$, $6$, $5$): at the highest gas velocities the detached material is carried by fewer but individually larger structures, some of which leave the region of interest, so $N_{\mathrm{det}}$ is a weaker indicator of atomisation than either $f_{\mathrm{det}}$ or the core penetration. What is monotone is the shortening of the intact core. The supplementary maps of Figure~\ref{fig:topology_supplementary} show that this is systematic across the envelope. Total interfacial density $\widehat{a}_{\mathrm{2D}}^{\,\mathrm{total}}$ (panel a) is comparatively flat, while the core contribution $\widehat{a}_{\mathrm{2D}}^{\,\mathrm{core}}$ (panel b) and the normalised core penetration $L_{\mathrm{core}}/L_{\mathrm{ROI}}$ (panel d) both fall steeply as $\Ug$ increases and $\Ul$ decreases.

\emph{Atomisation-aware design maps.} Figure~\ref{fig:design_maps} maps the quantities that follow from this decomposition. Panel (a) shows the detached interfacial density $\widehat{a}_{\mathrm{2D}}^{\,\mathrm{det}}$, which is largest in a band near $\Ul\approx3$ at high $\Ug$ where the column is thin enough to be stripped efficiently yet still supplied with liquid. Panel (b) shows the detached area fraction $A_{\mathrm{det}}/A_{\mathrm{liq}}$, which organises the envelope along a diagonal: fragmentation is promoted by high gas velocity and suppressed by high liquid velocity, consistent with the momentum-flux ratio controlling breakup. The dotted atomisation-onset contour follows that diagonal, and the region above and to its left, high $\Ul$ at low $\Ug$, fails the criterion of Equation~\eqref{eq:atomisation_onset} and is shown shaded in panel (c). Because that criterion requires only a single detached structure carrying $0.2\%$ of the liquid area, the shaded region marks conditions at which the predicted column does not detach anything at all; it is a floor on atomisation, not a design-quality threshold, and points just inside the feasible region should not be read as well atomised.

Panel (c) maps the constrained efficiency $\eta^*_{\mathrm{atom}}$ with the feasible ridge marked, and it revises the unconstrained picture materially. Judged on total interfacial area alone, efficiency appears maximised along the low-$\Ug$ edge, simply because $P^*_{\mathrm{gas}}$ grows cubically while total area does not. That optimum is an artefact of counting the intact column: at low gas velocity most of the liquid remains in an attached column, so the interfacial area credited there is largely unavailable for dispersed-phase transfer. Once the feasibility constraint is imposed, the ridge separates into two branches. Below $\Ul\approx4$, where breakup occurs even at modest gas velocity, the ridge remains at the low-$\Ug$ boundary. Above $\Ul\approx6$ the low-$\Ug$ region becomes infeasible and the ridge moves into the interior of the envelope, to $\Ug\approx55$--$70$. The efficient operating locus for a spray that must actually atomise is therefore an interior ridge rather than a boundary, which removes the concern that the optimum was an artefact of the edge of the trained range. It does not, however, remove the interpolation artefact: the ridge lies between the trained gas velocities $\Ug=50$, $60$, and $70$ in the exact internodal region, where the predicted area is biased low due to the smoothing of interpolated histories. The ridge should therefore be confirmed by evaluating the surrogate at the training nodes themselves and, ultimately, by CFD, before its position is taken as physical.

\begin{figure}[!htbp]
    \centering
    \includegraphics[width=0.8\linewidth]{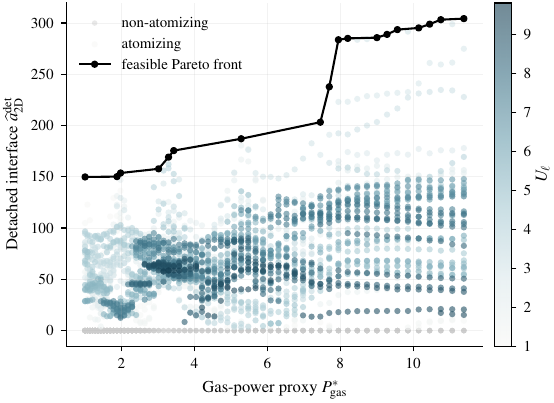}
    \caption{
    Detached interfacial density against the gas-power proxy $P^*_{\mathrm{gas}}$ for every swept operating point, coloured by liquid
    velocity $\Ul$. Points failing the atomisation criterion are shown separately
    and carry $\widehat{a}_{\mathrm{2D}}^{\,\mathrm{det}}\approx0$; the front is computed over the feasible set only. An order-of-magnitude increase in
    atomising power roughly doubles the attainable detached area, and the step near $P^*_{\mathrm{gas}}\approx7.9$ marks entry into the high-fragmentation band of Figure~\ref{fig:design_maps}(a).}
    \label{fig:pareto}
\end{figure}

\emph{Pareto front.} Figure~\ref{fig:pareto} recasts the sweep as a trade-off between detached interfacial area and gas power, with non-atomising points shown separately at $\widehat{a}_{\mathrm{2D}}^{\,\mathrm{det}}\approx0$. The feasible front rises from about $150$ at $P^*_{\mathrm{gas}}\approx1$ to about $305$ at $P^*_{\mathrm{gas}}\approx11.5$, so an order-of-magnitude increase in atomising power roughly doubles the attainable detached interfacial area. This is a steeper return than the total-area front suggests because, at low power, much of the total area is locked in the intact core and does not count. The front is not smooth: a pronounced step near $P^*_{\mathrm{gas}}\approx7.9$ carries the attainable area from about $203$ to about $285$, marking entry into the high-fragmentation band of Figure~\ref{fig:design_maps}(a). A step of this kind is the practically useful feature of the map, since it identifies a threshold in atomising power beyond which a disproportionate gain in dispersed area becomes available.

The engineering conclusion is a screening result rather than a final design. Over this envelope, on this geometry, and under a cubic gas-power cost, the efficient locus for atomising operation is an interior ridge at $\Ug\approx55$--$70$ for liquid velocities above $\Ul\approx6$, dropping to the low-$\Ug$ boundary only where the liquid throughput is small enough to break up unaided. A threshold near $P^*_{\mathrm{gas}}\approx7.9$ marks a step change in attainable dispersed area. Four qualifications bound it. The maps are evaluated at $\Dn=10$, where the interfacial-area bias is largest, and the rank correlation falls to $0.73$. The topology decomposition is applied to predicted fields whose small-scale structures have already been smoothed. It additionally discards unattached components below eight pixels, so $N_{\mathrm{det}}$ and $f_{\mathrm{det}}$ are lower bounds on the true fragmentation, and the finest droplets are invisible to it by construction; the atomisation-onset boundary is correspondingly conservative, and no operating point near that boundary has been checked against the solver. The ridge and the front are both read off interpolated histories, so the internodal bias described above is a candidate explanation for structure that lies between training nodes. Most importantly, the validation of Figure~\ref{fig:area_validation} establishes rank fidelity for \emph{total} interfacial area, predominantly on conditions the operator was trained on; the decomposed quantities $\widehat{a}_{\mathrm{2D}}^{\,\mathrm{det}}$ and $f_{\mathrm{det}}$ inherit that validation only indirectly and have not themselves been checked case by case against CFD. Confirming the onset boundary against the solver is the first thing a user of these maps should do.

\FloatBarrier

\section{Conclusion}

This work examined combinations of state representation, neural architecture, and physical regularisation and their adoption to multiphase-spray forecasting. The principal model we propose is a boundary-conditioned FNO that enables reconstruction of a phase field via a differentiable SDF-to-phase transformation. We added a physics-informed extension by including a target-increment balance, a narrowband Eikonal consistency, and phase boundedness while keeping the operator architecture fixed.
The principal findings are:
\begin{itemize}
    \item the signed-distance representation substantially improves FNO forecasting over direct phase-fraction prediction, reducing phase-field error by roughly a factor of three at $\Dn=10$, with the two models scored in their respective target spaces rather than against a common target;
    \item a direct-phase U-Net gives the lowest predictive error on this fixed-grid benchmark, on parameter counts that are not matched to the operator's;
    \item the inventory penalty is what secures liquid-content preservation: removing it raises the relative inventory error roughly fivefold at $\Dn=1$, at a small gain in signed-distance error, which makes the proposed model a deliberate compromise between local fidelity and global consistency;
    \item the signed-distance FNO preserves the dominant spray trajectory when extrapolated to $\Ug=90$, while fine structures are progressively smoothed;
    \item the physics-informed extension optimises stably and leaves phase-proxy accuracy, interface overlap, and liquid-inventory behaviour separated from the data-driven baseline by less than the case-to-case scatter, so its geometric and boundedness regularity comes at no predictive cost and at no predictive gain either, and its conservation term remains supervised by the CFD labels rather than enforced through boundary fluxes;
    \item interfacial area extracted from the predicted zero level set is under-predicted in absolute terms but rank-preserving across operating points ($\rho=0.93$ at $\Dn=1$, falling to $0.73$ at $\Dn=10$), which makes the surrogate a valid screening tool for the operating envelope of the injector it was trained on;
    \item the surrogate interrogates an additional operating point about $3.3\times10^{4}$ times faster than the reference solver, reducing a $3000$-point envelope sweep from roughly $67$ days of solver time to under three minutes, once the $27$ hours of training simulations have been paid for;
    \item considering only the detached interface and excluding non-atomising conditions moves the efficient operating locus off the low-$\Ug$ boundary into the interior of the envelope ($\Ug\approx55$--$70$) for liquid velocities above $\Ul\approx6$, and exposes a step in the feasible Pareto front near $P^*_{\mathrm{gas}}\approx7.9$ beyond which a disproportionate gain in dispersed interfacial area becomes available; both features lie between training nodes and should be confirmed at the nodes and against CFD before being read as physical.
\end{itemize}

The framework provides a structured benchmark for full-field multiphase forecasting and a demonstrated route from field-level prediction to injector design screening. The natural next step is to extend the framework to three-dimensional sprays to provide more reliable predictions. Another important extension will be to implement a physics-informed neural operator that uses multi-resolution constraints as well as pressure and velocity fields to enable more accurate and consistent interface tracking, and to test the mesh-independence that motivates the operator formulation by evaluating a trained model at resolutions other than the training grid. 

\section*{Data and code availability}
The CFD dataset generated for this study, the trained model weights, and the training and evaluation scripts are available from the corresponding author (P. Guida) upon reasonable request.

\section*{Funding}
This work was supported by King Abdullah University of Science and Technology (KAUST).

\section*{Conflicts of interest}
The authors declare that they have no known competing financial interests or personal relationships that could have appeared to influence the work reported in this paper.

\bibliographystyle{unsrt}
\bibliography{references}

@article{guida2026data,
  title={Data-driven prediction of interface evolution in multiphase flows using Fourier Neural Operators},
  author={Guida, Paolo and Roberts, William L},
  journal={Energy and AI},
  pages={100775},
  year={2026},
  publisher={Elsevier}
}

@inproceedings{ronneberger2015u,
  title={U-net: Convolutional networks for biomedical image segmentation},
  author={Ronneberger, Olaf and Fischer, Philipp and Brox, Thomas},
  booktitle={International Conference on Medical image computing and computer-assisted intervention},
  pages={234--241},
  year={2015},
  organization={Springer}
}

@book{lefebvre2017atomization,
  title={Atomization and sprays},
  author={Lefebvre, Arthur H and McDonell, Vincent G},
  year={2017},
  publisher={CRC press}
}

@article{fansler2015spray,
  title={Spray measurement technology: a review},
  author={Fansler, Todd D and Parrish, Scott E},
  journal={Measurement Science and Technology},
  volume={26},
  number={1},
  pages={012002},
  year={2015},
  publisher={IOP Publishing}
}

@article{xia2020experimental,
  title={Experimental study of injection characteristics under diesel’s sub/trans/supercritical conditions with various nozzle diameters and injection pressures},
  author={Xia, Jin and Zhang, Qiankun and Huang, Zhong and Ju, Dehao and Lu, Xingcai},
  journal={Energy conversion and management},
  volume={215},
  pages={112949},
  year={2020},
  publisher={Elsevier}
}

@article{rigas2016spray,
  title={Spray printing of organic semiconducting single crystals},
  author={Rigas, Grigorios-Panagiotis and Payne, Marcia M and Anthony, John E and Horton, Peter N and Castro, Fernando A and Shkunov, Maxim},
  journal={Nature Communications},
  volume={7},
  number={1},
  pages={13531},
  year={2016},
  publisher={Nature Publishing Group UK London}
}

@article{shrestha2023primary,
  title={Primary spray breakup from a nasal spray atomizer using volume of fluid to discrete phase model},
  author={Shrestha, Kendra and Van Strien, James and Fletcher, David F and Inthavong, Kiao},
  journal={Physics of Fluids},
  volume={35},
  number={5},
  year={2023},
  publisher={AIP Publishing}
}

@article{di2022computational,
  title={Computational fluid dynamics characterization of the hollow-cone atomization: Newtonian and non-Newtonian spray comparison},
  author={Di Martino, Massimiliano and Ahirwal, Deepak and Maffettone, Pier Luca},
  journal={Physics of Fluids},
  volume={34},
  number={9},
  year={2022},
  publisher={AIP Publishing}
}

@article{moghimi2025simulations,
  title={Simulations of sprays in vessels and the effects of key variables for understanding cleaning processes},
  author={Moghimi, Mohsen H and Karimi-Jafari, Maryam and Shardt, Orest},
  journal={Chemical Engineering Research and Design},
  volume={217},
  pages={128--151},
  year={2025},
  publisher={Elsevier}
}

@article{sun2024liquid,
  title={Liquid sheet formation and spray characterization of N-heptane spray jet from a swirl atomizer: Numerical analysis and validation},
  author={Sun, Yaquan and Vegad, Chetankumar S and Li, Yongxiang and Dre{\ss}ler, Louis and Renou, Bruno and Nishad, Kaushal and Demoulin, Fran{\c{c}}ois-Xavier and Hasse, Christian and Sadiki, Amsini},
  journal={Physics of Fluids},
  volume={36},
  number={3},
  year={2024},
  publisher={AIP Publishing}
}

@article{azizzadenesheli2024neural,
  title={Neural operators for accelerating scientific simulations and design},
  author={Azizzadenesheli, Kamyar and Kovachki, Nikola and Li, Zongyi and Liu-Schiaffini, Miguel and Kossaifi, Jean and Anandkumar, Anima},
  journal={Nature Reviews Physics},
  volume={6},
  number={5},
  pages={320--328},
  year={2024},
  publisher={Nature Publishing Group UK London}
}

@article{lin2026operator,
  title={Operator learning augmented physics-informed neural networks for partial differential equations exhibiting sharp features},
  author={Lin, Bin and Mao, Zhiping and Wang, Zhicheng},
  journal={Physical Review E},
  volume={113},
  number={3},
  pages={035306},
  year={2026},
  publisher={APS}
}

@article{heinrich20203d,
  title={3D-coupling of Volume-of-Fluid and Lagrangian particle tracking for spray atomization simulation in OpenFOAM},
  author={Heinrich, Martin and Schwarze, R{\"u}diger},
  journal={SoftwareX},
  volume={11},
  pages={100483},
  year={2020},
  publisher={Elsevier}
}

@article{ramlau2026learning,
  title={Learning Interface Breakup: A Geometry-Conditioned Latent Surrogate for Spray Formation},
  author={Ramlau, Julius H and Hastedt, Friedrich and Birdal, Tolga and Chanona, Ehecatl-Antonio del R{\'\i}o and Basha, Nausheen S and Matar, Omar K},
  journal={arXiv preprint arXiv:2606.16587},
  year={2026}
}

@inproceedings{rahman2016optimizing,
  title={Optimizing intersection-over-union in deep neural networks for image segmentation},
  author={Rahman, Md Atiqur and Wang, Yang},
  booktitle={International symposium on visual computing},
  pages={234--244},
  year={2016},
  organization={Springer}
}

@article{salehi2025data,
  title={Data-driven modelling of spray flows: Current status and future direction},
  author={Salehi, Fatemeh and Beheshti, Amin and Eftekharian, Esmaeel and Chen, Longfei and Hardalupas, Yannis},
  journal={Journal of the Energy Institute},
  volume={119},
  pages={101991},
  year={2025},
  publisher={Elsevier}
}

@article{yariv2021volume,
  title={Volume rendering of neural implicit surfaces},
  author={Yariv, Lior and Gu, Jiatao and Kasten, Yoni and Lipman, Yaron},
  journal={Advances in neural information processing systems},
  volume={34},
  pages={4805--4815},
  year={2021}
}

@article{gropp2020implicit,
  title={Implicit geometric regularization for learning shapes},
  author={Gropp, Amos and Yariv, Lior and Haim, Niv and Atzmon, Matan and Lipman, Yaron},
  journal={arXiv preprint arXiv:2002.10099},
  year={2020}
}

@article{dong2023development,
  title={Development of machine learning based droplet diameter prediction model for electrohydrodynamic atomization systems},
  author={Dong, Tao and Wang, Jin-Xin and Wang, Yong and Tang, Guan-Hua and Cheng, Yongpan and Yan, Wei-Cheng},
  journal={Chemical Engineering Science},
  volume={268},
  pages={118398},
  year={2023},
  publisher={Elsevier}
}

@book{ishii2010thermo,
  title={Thermo-fluid dynamics of two-phase flow},
  author={Ishii, Mamoru and Hibiki, Takashi},
  year={2010},
  publisher={Springer Science \& Business Media}
}

@article{desjardins2008accurate,
  title={An accurate conservative level set/ghost fluid method for simulating turbulent atomization},
  author={Desjardins, Olivier and Moureau, Vincent and Pitsch, Heinz},
  journal={Journal of Computational Physics},
  volume={227},
  number={18},
  pages={8395--8416},
  year={2008},
  publisher={Elsevier}
}

@article{li2020fourier,
  title={Fourier neural operator for parametric partial differential equations},
  author={Li, Zongyi and Kovachki, Nikola and Azizzadenesheli, Kamyar and Liu, Burigede and Bhattacharya, Kaushik and Stuart, Andrew M and Anandkumar, Anima},
  journal={arXiv preprint arXiv:2010.08895},
  year={2020}
}

@article{tran2021factorized,
  title={Factorized fourier neural operators},
  author={Tran, Alasdair and Mathews, Alexander and Xie, Lexing and Ong, Cheng Soon},
  journal={arXiv preprint arXiv:2111.13802},
  year={2021}
}

@inproceedings{NEURIPS2023_940a7634,
 author = {Liu, Ning and Jafarzadeh, Siavash and Yu, Yue},
 booktitle = {Advances in Neural Information Processing Systems},
 editor = {A. Oh and T. Naumann and A. Globerson and K. Saenko and M. Hardt and S. Levine},
 pages = {47438--47450},
 publisher = {Curran Associates, Inc.},
 title = {Domain Agnostic Fourier Neural Operators},
 url = {https://proceedings.neurips.cc/paper_files/paper/2023/file/940a7634dab556b67af15bacd337f7db-Paper-Conference.pdf},
 volume = {36},
 year = {2023}
}

@article{duruisseaux2025fourier,
  title={Fourier Neural Operators Explained: A Practical Perspective},
  author={Duruisseaux, Valentin and Kossaifi, Jean and Anandkumar, Anima},
  journal={arXiv preprint arXiv:2512.01421},
  year={2025}
}

@incollection{goswami2023physics,
  title={Physics-informed deep neural operator networks},
  author={Goswami, Somdatta and Bora, Aniruddha and Yu, Yue and Karniadakis, George Em},
  booktitle={Machine learning in modeling and simulation: methods and applications},
  pages={219--254},
  year={2023},
  publisher={Springer}
}

@article{lyu2023multi,
  title={Multi-fidelity prediction of fluid flow based on transfer learning using Fourier neural operator},
  author={Lyu, Yanfang and Zhao, Xiaoyu and Gong, Zhiqiang and Kang, Xiao and Yao, Wen},
  journal={Physics of Fluids},
  volume={35},
  number={7},
  year={2023},
  publisher={AIP Publishing}
}

@article{zhang2022fourier,
  title={Fourier neural operator for solving subsurface oil/water two-phase flow partial differential equation},
  author={Zhang, Kai and Zuo, Yuande and Zhao, Hanjun and Ma, Xiaopeng and Gu, Jianwei and Wang, Jian and Yang, Yongfei and Yao, Chuanjin and Yao, Jun},
  journal={Spe Journal},
  volume={27},
  number={03},
  pages={1815--1830},
  year={2022},
  publisher={OnePetro}
}

@article{jain2025scaling,
  title={Scaling the predictions of multiphase flow through porous media using operator learning},
  author={Jain, Navya and Roy, Shantanu and Kodamana, Hariprasad and Nair, Prapanch},
  journal={Chemical Engineering Journal},
  volume={503},
  pages={157671},
  year={2025},
  publisher={Elsevier}
}

@article{ma2024enhancing,
  title={Enhancing subsurface multiphase flow simulation with Fourier neural operator},
  author={Ma, Xianlin and Zhong, Rong and Zhan, Jie and Zhou, Desheng},
  journal={Heliyon},
  volume={10},
  number={18},
  year={2024},
  publisher={Elsevier}
}

@article{dong2025multi,
  title={Multi-scale enhanced multiwavelet-based operator learning model for multiphase flow simulation},
  author={Dong, Yunlong and Song, Tao and Li, Xue and Han, Peifu and Zhao, Peizhi and Zhai, Chuchu and Jing, Fengrui and Hao, Long},
  journal={Physics of Fluids},
  volume={37},
  number={3},
  year={2025},
  publisher={AIP Publishing}
}

@article{qiu2022physics,
  title={Physics-informed neural networks for phase-field method in two-phase flow},
  author={Qiu, Rundi and Huang, Renfang and Xiao, Yao and Wang, Jingzhu and Zhang, Zhen and Yue, Jieshun and Zeng, Zhong and Wang, Yiwei},
  journal={Physics of Fluids},
  volume={34},
  number={5},
  year={2022},
  publisher={AIP Publishing}
}

@article{hao2024fourier,
  title={Fourier neural operator networks for solving reaction--diffusion equations},
  author={Hao, Yaobin and Song, Fangying},
  journal={Fluids},
  volume={9},
  number={11},
  pages={258},
  year={2024},
  publisher={MDPI}
}

@article{guida2022computational,
  title={A computational study of thermally induced secondary atomization in multicomponent droplets},
  author={Guida, Paolo and Ceschin, Alberto and Saxena, Saumitra and Im, Hong G and Roberts, William L},
  journal={Journal of Fluid Mechanics},
  volume={935},
  year={2022},
  publisher={Cambridge University Press}
}

@article{GAMET2020104722,
title = "Validation of volume-of-fluid OpenFOAM® isoAdvector solvers using single bubble benchmarks",
journal = "Computers \& Fluids",
volume = "213",
pages = "104722",
year = "2020",
issn = "0045-7930",
doi = "https://doi.org/10.1016/j.compfluid.2020.104722",
author = "Lionel Gamet and Marco Scala and Johan Roenby and Henning Scheufler and Jean-Lou Pierson"
}

@inproceedings{roenby2017new,
  title={A new volume-of-fluid method in OpenFOAM},
  author={Roenby, Johan and Larsen, Bjarke Eltard and Bredmose, Henrik and Jasak, Hrvoje},
  booktitle={Marine vi: Proceedings of the vi international conference on computational methods in marine engineering},
  pages={266--277},
  year={2017},
  organization={CIMNE}
}

@article{hirt1981volume,
  title={Volume of fluid ({VOF}) method for the dynamics of free boundaries},
  author={Hirt, Cyril W and Nichols, Billy D},
  journal={J. Comput. Phys.},
  volume={39},
  number={1},
  pages={201--225},
  year={1981},
  publisher={Elsevier}
}

@article{raissi2019physics,
  title={Physics-informed neural networks: A deep learning framework for solving forward and inverse problems involving nonlinear PDEs},
  author={Raissi, Maziar and Perdikaris, Paris and Karniadakis, George E},
  journal={J. Comput. Phys.},
  volume={378},
  pages={686--707},
  year={2019}
}

@article{scheufler2019accurate,
  title={Accurate and efficient surface reconstruction from volume fraction data on general meshes},
  author={Scheufler, Henning and Roenby, Johan},
  journal={J. Comput. Phys.},
  volume={383},
  pages={1--23},
  year={2019},
  publisher={Elsevier}
}

@article{scheufler2023twophaseflow,
  title={TwoPhaseFlow: A framework for developing two phase flow solvers in OpenFOAM},
  author={Scheufler, Henning and Roenby, Johan},
  journal={OpenFOAM{\textregistered} Journal},
  volume={3},
  pages={200--224},
  year={2023}
}

@article{chen2026large,
  title={Large-eddy simulations of spray dynamics in an air-assisted atomizer using a geometric volume-of-fluid method},
  author={Chen, Po-Han and Ceschin, Alberto and Shakeel, Mohammad Raghib and Im, Hong G},
  journal={International Journal of Spray and Combustion Dynamics},
  volume={18},
  number={2},
  pages={86--104},
  year={2026},
  publisher={SAGE Publications Sage UK: London, England}
}

@article{chen2026effects,
  title={Effects of recess on the spray dynamics in coaxial air-assisted atomizers},
  author={Chen, Po-Han and Ceschin, Alberto and Shakeel, Mohammad Raghib and Im, Hong G},
  journal={Proceedings of the Combustion Institute},
  volume={42},
  pages={106246},
  year={2026},
  publisher={Elsevier}
}

@article{zhang2024physics,
  title={Physics-informed neural networks for multiphase flow in porous media considering dual shocks and interphase solubility},
  author={Zhang, Jingjing and Braga-Neto, Ulisses and Gildin, Eduardo},
  journal={Energy \& Fuels},
  volume={38},
  number={18},
  pages={17781--17795},
  year={2024},
  publisher={ACS Publications}
}

@article{gamet2020validation,
  title={Validation of volume-of-fluid OpenFOAM{\textregistered} isoAdvector solvers using single bubble benchmarks},
  author={Gamet, Lionel and Scala, Marco and Roenby, Johan and Scheufler, Henning and Pierson, Jean-Lou},
  journal={Computers \& Fluids},
  volume={213},
  pages={104722},
  year={2020},
  publisher={Elsevier}
}

@article{cummins2005estimating,
  title={Estimating curvature from volume fractions},
  author={Cummins, Sharen J and Francois, Marianne M and Kothe, Douglas B},
  journal={Computers \& structures},
  volume={83},
  number={6-7},
  pages={425--434},
  year={2005},
  publisher={Elsevier}
}

\end{document}